\documentclass[12pt,thmsa]{article}
\usepackage{amsfonts}

\input{tcilatex}
\begin{document}

\author{\textbf{Senya Shlosman}\thanks{%
CPT, Luminy, Marseille and IITP, Russian Academy of Sciences. The work was
partially supported by the NSF through grant DMS 9800860 and by the Russian
Fund for Fundamental research through grant 99-01-00284. Email :
shlosman@cpt.univ-mrs.fr.}}
\title{\textbf{Geometric variational problems of statistical mechanics and of
combinatorics}}
\date{\textbf{May, 17, 1999}}
\maketitle

\begin{abstract}
We present the geometric solutions of the various extremal problems of
statistical mechanics and combinatorics. Together with the Wulff
construction, which predicts the shape of the crystals, we discuss the
construction which exhibit the shape of a typical Young diagram and of a
typical skyscraper.
\end{abstract}

\section{\protect\smallskip Introduction}

\subsection{Statistical mechanics}

The variational problems of statistical mechanics we are going to discuss
here are those related to the formation of a droplet or a crystal of one
substance inside another. The question here is: what shape such a formation
would take? The statement that such shape should be defined by the minimum
of the overall surface energy subject to the volume constraint was known
from the times immemorial. In the isotropic case, when the surface tension
does not depend on the orientation of the surface, and so is just a positive
number, the shape in question should be of course spherical (provided we
neglect the gravitational effects). In a more general situation the shape in
question is less symmetric. The corresponding variational problem is called
the \textit{Wulff problem}. Wulff formulated it in his paper \cite{W} of
1901, where he also presented a geometric solution to it, called the \textit{%
Wulff construction} (see section 2.2 below).

This Wulff construction was considered by the rigorous statistical mechanics
as just a phenomenological statement, though the notion of the surface
tension was among its central notions. The situation changed after the
appearance of the book \cite{DKS}. There it was shown that in the setting of
the canonical ensemble formalism, in the regime of the first order phase
transition, the (random) shape occupied by one of the phases has
asymptotically (in the thermodynamic limit) a non-random shape, given
precisely by the Wulff construction! In other words, a typical macroscopic
random droplet looks very close to the Wulff shape. The results of the book 
\cite{DKS} are restricted to the 2D Ising ferromagnet at low temperature,
though the methods of the book are suitable for the rigorous treatment of
much more general two-dimensional low-temperature models. Physical intuition
is that as soon as there is phase coexistence, these results should be
valid. It was proven in \cite{I1, I2, IS} to be the case for the 2D Ising
model at all subcritical temperatures. Some results for the higher
dimensional case were obtained in \cite{Bo,CeP}. For the independent
percolation the corresponding results were obtained in \cite{ACC} for the 2D
case, and in \cite{Ce} in the 3D case.

\subsection{Combinatorics.}

The main content of the present paper concerns the problems arising in
combinatorics, so in this section we describe some of them in more details.

A partition $p$ of an integer $N$ is a collection of non-negative integers $%
n_1\ge n_2\ge ...\ge n_k\ge ...,$ such that $\sum_{i=1}^\infty n_i=N.$ It
can be specified by the sequence $\left\{ r_k\right\} $ of integers, with $%
r_k=l$ iff exactly $l$ elements of $p$ equal $k.$ It can also be described
by the monotone function 
\[
\phi _p\left( y\right) =\sum_{k=\left\lceil y\right\rceil }^\infty r_k. 
\]
Its graph $G\left[ \phi _p\right] $ provides a graphical description of $p$
and is called a \textit{(2D)} \textit{Young diagram}.

Similarly, a plane partition $P$ of an integer $N$ is a two-dimensional
array of non-negative integers $n_{ij},$ such that for any $i$ we have $%
n_{i1}\ge n_{i2}\ge ...\ge n_{ik}\ge ...,$ for any $j$ we have $n_{1j}\ge
n_{2j}\ge ...\ge n_{kj}\ge ...,$ while again $\sum_{i,j=1}^\infty n_{ij}=N.$
One defines the corresponding function $\phi _P\left( y_1,y_2\right) $ in
the obvious way. The function $\phi _P\left( y_1,y_2\right) $ is monotone in
each variable. Its graph $G\left[ \phi _P\right] $ is called a \textit{3D} 
\textit{Young diagram} or a \textit{skyscraper}.

Many more objects of a similar type can be defined. For example, one can put
restrictions on how the steps of the stair $G\left[ \phi _p\right] $ can
look: they can not be longer than 3 units, and their heights can be only 1,2
or 5, say. The same freedom is allowed in 3D, and above.

Let us fix the number $N,$ choose the kind of diagrams we are interested in,
and consider the corresponding set $\mathcal{D}_N$ of all these diagrams.
There are finitely many of them, so we can put a uniform probability
distribution on $\mathcal{D}_N.$ (Here, again, variations are possible.) The
question now is the following: how the typical diagram from the family $%
\mathcal{D}_N$ looks like, when $N\rightarrow \infty ?$

The first problem of that type was solved in the paper \cite{VK}, see also 
\cite{V1,V2, DVZ}. It was found there, that the typical 2D Young diagram
under statistics described above, if scaled by the factor $\left( 1/\sqrt{N}%
\right) ,$ tends to the curve 
\begin{equation}
\exp \left\{ -\dfrac \pi {\sqrt{6}}x\right\} +\exp \left\{ -\dfrac \pi {%
\sqrt{6}}y\right\} =1.  \label{08}
\end{equation}
More precisely, for every $\varepsilon >0$ the probability that the scaled
Young diagram would be within distance $\varepsilon $ from the curve (\ref
{08}), goes to $1$ as $N\rightarrow \infty .$

The heuristic way to obtain (\ref{08}) (and similar results) is the
following:

$i)$ Let $A=\left( a_1,a_2\right) ,B=\left( b_1,b_2\right) $ be two points
in $\Bbb{Z}^2$, with $a_1<b_1,a_2>b_2.$ We can easily see that the number $%
\#\left( A,B\right) $ of lattice staircases, starting from $A$, terminating
at $B,$ and allowed to go only to the right or down, is given by $\binom{%
\left( b_1-a_1\right) +\left( a_2-b_2\right) }{\left( b_1-a_1\right) }.$
Therefore one concludes by using the Stirling formula that 
\begin{equation}
\lim_{\left| B-A\right| \rightarrow \infty }\frac 1{\left| B-A\right| }\ln
\#\left( A,B\right) =h\left( \mathbf{n}_{AB}\right) .  \label{24}
\end{equation}
Here $\mathbf{n}_{AB}$ is the unit vector, normal to the segment $\left[
A,B\right] ,$ and for $\mathbf{n}=\left( n_1,n_2\right) ,$ $\alpha =\frac{n_1%
}{n_1+n_2},$ the \textit{entropy function} $h\left( \mathbf{n}\right)
=-\left( \alpha \ln \alpha +\left( 1-\alpha \right) \ln \left( 1-\alpha
\right) \right) .$

$ii)$ One argues that the number of Young diagrams of the area $N$ scaled by 
$\sqrt{N},$ ``going along'' the monotone curve $y=c\left( x\right) \ge 0$
with integral one, is approximately given by 
\begin{equation}
\exp \left\{ \sqrt{N}\int h\left( -\frac{c^{\prime }\left( x\right) }{\sqrt{%
1+\left( c^{\prime }\left( x\right) \right) ^2}},\frac 1{\sqrt{1+\left(
c^{\prime }\left( x\right) \right) ^2}}\right) \sqrt{1+\left( c^{\prime
}\left( x\right) \right) ^2}dx\right\} .  \label{17}
\end{equation}

$iii)$ Assuming that indeed the model under consideration exhibits under a
proper scaling some typical behavior, described by a nice smooth non-random
curve (or surface) $\mathcal{C},$ one comes to the conclusion that the curve 
$\mathcal{C}$ should be such that the integral in (\ref{17}), computed along 
$\mathcal{C},$ is maximal compared with all other allowed curves.

In general case one is not able to write down the corresponding entropy
function precisely. The only information available generally is the
existence of the limit of the type of (\ref{24}), by a subadditivity
argument. It should be stressed that even when the variational problem for
the model is known, the main difficulty of the rigorous treatment of the
model is the proof that indeed it does exhibit a nontrivial behavior after a
proper scaling.

The above program was realized in \cite{V1,V2}, see also \cite{DVZ}, for the
2D case described above and for some other cases. In \cite{Bl} a class of
more general 2D problems was studied. The first 3D problem was successfully
studied in \cite{CKP}. The method of the last paper can also solve the
skyscraper problem, as is claimed in \cite{Ke}.

When compared with the situation in statistical mechanics, the combinatorial
program and its development look very similar. The only difference is that
the counterpart of the Wulff construction was not designed in combinatorics,
probably because there was no heuristic period there. In this note we fill
this lack of parallelism by presenting such a construction. It provides,
like the Wulff one, the geometric solution to the corresponding variational
problem under minimal restrictions on the initial data, and also proves the
uniqueness of the solution.

In the next section we first remind the reader about the Wulff minimizing
problem (sect. 2.1) and the Wulff construction (sect. 2.2), which solves
this problem, and then present the corresponding maximizing problem of
combinatorics (sect. 2.3) and the geometric construction for its solution
(sect 2.4), which is our main result. We give the proof in the section 3.

\section{Statement of results}

\subsection{Wulff minimizing problem.}

Let $S^d\subset \Bbb{R}^{d+1}$ denote the unit sphere, and let the real
function $\tau $ on $S^d$ be given. We suppose that the function is
continuous, positive: $\tau \left( \mathbf{\cdot }\right) \ge const>0,$ and
even: $\tau \left( \mathbf{n}\right) =\tau \left( -\mathbf{n}\right) .$ Then
for every hypersurface $M^d\subset \Bbb{R}^{d+1}$ we can define the \textit{%
Wulff functional} 
\begin{equation}
\mathcal{W}_\tau \left( M^d\right) =\int_{M^d}\tau \left( \mathbf{n}%
_x\right) \,ds_x.  \label{01}
\end{equation}
Here $x\in M^d$ is a point on the manifold $M^d,$ the vector $\mathbf{n}_x$
is the unit vector parallel to the normal to $M^d$ at $x,$ and $ds$ is the
usual volume $d$-form on $M^d,$ induced from the Riemannian metric on $\Bbb{R%
}^{d+1}$ by the embedding $M^d\subset \Bbb{R}^{d+1}.$ Of course, we need to
assume that the normal to $M^d$ is defined almost everywhere, i.e. that $M^d$
is smooth enough. Let now $D_q$ be the collection of all \textit{closed}
hypersurfaces $M^d$, \textit{embedded }in $\Bbb{R}^{d+1},$ and such that the
volume $\mathrm{vol}\left( M^d\right) $ inside\textit{\ }$M^d$ equals $q.$
The \textit{Wulff problem }consists in finding the lower bound of $\mathcal{W%
}_\tau $ over $D_1:$%
\begin{equation}
w_\tau =\inf_{M\in D_1}\mathcal{W}_\tau \left( M\right) ,  \label{02}
\end{equation}
as well as the minimizing surface(s) $W_\tau ,$ such that $\mathcal{W}_\tau
\left( W_\tau \right) =w_\tau ,$ if it exists. It turns out that the above
variational problem indeed can be solved. It has a unique solution, which is
given by the following

\subsection{Wulff construction (\protect\cite{W}).}

The minimizer $W_\tau $ can be obtained as follows. For every $\mathbf{n}\in
S^d,\,\lambda >0$ define the half-space 
\begin{equation}
L_\tau ^{<}\left( \mathbf{n;}\lambda \right) =\left\{ \mathbf{x}\in \Bbb{R}%
^{d+1}:\left( \mathbf{x},\mathbf{n}\right) \le \lambda \,\tau \left( \mathbf{%
n}\right) \right\} ,  \label{11}
\end{equation}
and let 
\begin{equation}
K_\tau ^{<}\left( \lambda \right) =\bigcap_{\mathbf{n}\in S^d}L_\tau
^{<}\left( \mathbf{n;}\lambda \right) ,  \label{12}
\end{equation}
\begin{equation}
M_\tau \left( \lambda \right) =\partial \left( K_\tau ^{<}\left( \lambda
\right) \right) .  \label{13}
\end{equation}
The bodies $K_\tau ^{<}\left( \lambda \right) $ are called \textit{Wulff
bodies. }We define $\lambda _1$ as the value of $\lambda ,$ for which $%
\mathrm{vol}\left( M_\tau \left( \lambda \right) \right) =1.$ Then we define 
$W_\tau =M_\tau \left( \lambda _1\right) .$ The surface $W_\tau $ is called
the \textit{Wulff shape}. This is the minimizer we are looking for.

The paper \cite{T2} contains a simple proof that $\mathcal{W}_\tau \left(
W_\tau \right) \le \mathcal{W}_\tau \left( M\right) $ for every $M\in D_1.$
The uniqueness of the minimizing surface is proven in \cite{T1}. It is known
that in dimension 2 the minimizing surface $W_\tau $ of the functional $%
\mathcal{W}_\tau $ is not only unique, but also is stable in the Hausdorf
metric; for the proof, see \cite{DKS}, Sect. 2.4.

\subsection{ Maximizing problem.}

In a dual problem we again have a function $\eta $ of a unit vector, but
this time it is defined only over the subset $\Delta ^d=$ $S^d\cap \Bbb{R}%
_{+}^{d+1}\ $of them, lying in the positive octant. We suppose again that
the function is continuous and nonnegative: $\eta \left( \mathbf{\cdot }%
\right) \ge 0.$ We assume additionally that 
\begin{equation}
\eta \left( \mathbf{n}\right) \rightarrow 0\text{ uniformly as }\mathbf{n}%
\rightarrow \partial \,\Delta ^d.  \label{05}
\end{equation}
Let now $G\subset \Bbb{R}_{+}^{d+1}$ be an embedded hypersurface. We assume
that for almost every $x\mathbf{\in }G$ the normal vector $\mathbf{n}_x$ is
defined, and moreover 
\begin{equation}
\mathbf{n}_x\in \Delta ^d\text{ for a.e. }x\mathbf{\in }G.  \label{07}
\end{equation}
Then we can define the functional 
\begin{equation}
\mathcal{V}_\eta \left( G\right) =\int_G\eta \left( \mathbf{n}_x\right)
\,ds_x.  \label{03}
\end{equation}
In analogy with the section 2.1 we introduce the families $\bar{D}_q,q>0,$
of such surfaces $G$ as follows:

$G\in \bar{D}_q$ iff

\noindent $\,i)$ $\,G$ splits the octant $\Bbb{R}_{+}^{d+1}$ into two parts,
with the boundary $\partial \Bbb{R}_{+}^{d+1}$ belonging to one of them,

\noindent $ii)$ the ($\left( d+1\right) $-dimensional) volume of the body $%
Q\left( G\right) ,$ enclosed between $\partial \Bbb{R}_{+}^{d+1}$ and $G,$
equals $q.$ In what follows we denote the volume of $Q\left( G\right) $ by $%
\mathrm{vol}\left( G\right) .$

\noindent For example, let $f\left( y\right) \ge 0$ be a function on $\Bbb{R}%
_{+}^d,$ non-increasing in each of $d$ variables, and $G\left[ f\right]
\subset $ $\Bbb{R}_{+}^{d+1}$ be its graph. Then 
\begin{equation}
\mathrm{vol}\left( G\left[ f\right] \right) =\int_{\Bbb{R}_{+}^d}f\left(
y\right) \,dy,  \label{06}
\end{equation}
so if $\int_{\Bbb{R}_{+}^d}f\left( y\right) \,dy=q,$ then $G\left[ f\right] $
is an element of $\bar{D}_q,$ provided the function $f$ is sufficiently
smooth.

Our problem now\textit{\ }is to find the \textit{upper bound} of $\mathcal{V}%
_\eta $ over $\bar{D}_1:$%
\begin{equation}
v_\eta =\sup_{G\in \bar{D}_1}\mathcal{V}_\eta \left( G\right) ,  \label{04}
\end{equation}
as well as the maximizing surface(s) $V_\eta \in \bar{D}_1,$ such that $%
\mathcal{V}_\eta \left( V_\eta \right) =v_\eta ,$ if possible. Note that the
last problem differs crucially from (\ref{02}), since here we are looking
for the \textit{supremum. }In particular, this upper bound evidently
diverges if taken over all surfaces, and not only over ``monotone'' one, in
the sense of (\ref{07}), unlike in the problem (\ref{02}).

It turns out that there exists a geometric construction, which provides a
solution to the variational problem (\ref{04}), in the same way as the Wulff
construction solves the problem (\ref{02}).

\subsection{The main result.}

For every $\mathbf{n}\in \Delta ^d,\,\lambda >0$ define the half-space 
\begin{equation}
L_\eta ^{>}\left( \mathbf{n;}\lambda \right) =\left\{ \mathbf{x}\in \Bbb{R}%
^{d+1}:\left( \mathbf{x},\mathbf{n}\right) \ge \lambda \,\eta \left( \mathbf{%
n}\right) \right\} ,  \label{14}
\end{equation}
and let 
\begin{equation}
K_\eta ^{>}\left( \lambda \right) =\bigcap_{\mathbf{n}\in \Delta ^d}L_\eta
^{>}\left( \mathbf{n;}\lambda \right)  \label{15}
\end{equation}
\begin{equation}
G_\eta \left( \lambda \right) =\partial \left( K_\eta ^{>}\left( \lambda
\right) \right) .  \label{16}
\end{equation}
Because of (\ref{05}), the surfaces $G_\eta \left( \lambda \right) $ are
graphs of functions, $f_\eta ^\lambda \left( y\right) ,\,y\in \Bbb{R}_{+}^d,$
i.e. $G_\eta \left( \lambda \right) =G\left[ f_\eta ^\lambda \right] .$

\begin{theorem}
Suppose the integrals $\mathrm{vol}\left( G_\eta \left( \lambda \right)
\right) $ (see (\ref{06})) are converging. Then the functional $\mathcal{V}%
_\eta $ has a unique maximizer, $V_\eta ,$ over the set $\bar{D}_1.$ It is
given by the above construction (\ref{16}): 
\[
V_\eta =\left( G_\eta \left( \lambda _1\right) \right) \equiv G\left[ f_\eta
^{\lambda _1}\right] ,
\]
where $\lambda _1$ satisfies $\mathrm{vol}\left( G_\eta \left( \lambda
_1\right) \right) =1,$ and the maximum of the functional $v_\eta =\mathcal{V}%
_\eta \left( V_\eta \right) ,$ (see (\ref{04})). If the integrals $\mathrm{%
vol}\left( G_\eta \left( \lambda \right) \right) $ diverge, then $v_\eta
=\infty .$
\end{theorem}

As we already said in the introduction, in all known cases the heuristic
arguments of the Section 1.2 turn out to be correct, and are validated by
corresponding (sometime quite hard) theorems proven. For example, they are
valid for the problem of finding the asymptotic shape of the Young diagram,
described in the Section 1.2, as was proven in \cite{VK,V1,V2}. Therefore,
the following statement holds:

\begin{corollary}
In the notations of the theorem above, the curve $\exp \left\{ -\frac \pi {%
\sqrt{6}}x\right\} +\exp \left\{ -\frac \pi {\sqrt{6}}y\right\} =1$ from the
formula (\ref{08}) coincides with the curve $G_h\left( \lambda _1\right) ,$
given by our construction applied to the function $\eta \left( \mathbf{n}%
\right) =h\left( \mathbf{n}\right) $ from the formula (\ref{24}).
\end{corollary}

Of course, this statement can also be easily checked directly.

\section{The proof of the Theorem.}

We start with the case of finite volumes: $\mathrm{vol}\left( G_\eta \left(
\lambda \right) \right) <\infty $ for all $\lambda .$

We will prove our theorem by showing that for any surface $G\in \bar{D}%
_1,G\neq V_\eta ,$ which coincides with $G\left[ V_\eta \right] $ outside
some big ball around the origin of $\Bbb{R}^{d+1},$ we have 
\[
\mathcal{V}_\eta \left( G\right) >\mathcal{V}_\eta \left( V_\eta \right) . 
\]

First, we need more detailed notation than in the previous section. For
every $\mathbf{x}\in $ $\Bbb{R}^{d+1},\mathbf{n}\in S^d,\kappa >0$ we define
the half-spaces 
\[
L^{>}\left( \mathbf{x,n;}\kappa \right) =\left\{ \mathbf{y}\in \Bbb{R}%
^{d+1}:\left( \mathbf{y}-\mathbf{x},\mathbf{n}\right) \ge \kappa \right\} 
\]
and the planes 
\[
L^{=}\left( \mathbf{x,n;}\kappa \right) =\left\{ \mathbf{y}\in \Bbb{R}%
^{d+1}:\left( \mathbf{y}-\mathbf{x},\mathbf{n}\right) =\kappa \right\} . 
\]
Let $C\subset \Bbb{R}^{d+1}$ be a convex set, and $\mathbf{x}\in C.$ The 
\textit{support function} $\tau _{\mathbf{x},C}\left( \cdot \right) $ is
defined by 
\[
\tau _{\mathbf{x},C}\left( \mathbf{n}\right) =\inf \left\{ \kappa
:L^{>}\left( \mathbf{x,n;}\kappa \right) \cap C=\emptyset \right\} ; 
\]
we put $\tau _{\mathbf{x},C}\left( \mathbf{n}\right) =\infty $ if $%
L^{>}\left( \mathbf{x,n;}\kappa \right) \cap C\neq \emptyset $ for all $%
\kappa .$ We denote by $\mathsf{K}$ the convex set $K_\eta ^{>}\left(
\lambda =\lambda _1\right) ,$ introduced in (\ref{15}), and we use the
notation $\mathsf{G}$ for the surface $\partial \left( \mathsf{K}\right) .$

Let $\varepsilon >0.$ Introduce the set $\Bbb{R}_\varepsilon ^{d+1}=\left\{ 
\mathbf{y}=\left( y_1,...,y_{d+1}\right) \in \Bbb{R}_{+}^{d+1}:y_i\ge
\varepsilon \right\} ,$ and define $\mathsf{K}\left( \varepsilon \right) =%
\mathsf{K}\cap \Bbb{R}_\varepsilon ^{d+1},\;\mathsf{G}\left( \varepsilon
\right) =\,\partial \left( \mathsf{K}\left( \varepsilon \right) \right) .$
The family of the subsets $\mathsf{\bar{G}}\left( \varepsilon \right) \equiv
\left( \mathsf{G}\left( \varepsilon \right) \cap \mathsf{G}\right) \subset 
\mathsf{G}$ is increasing, with $\cup _{\varepsilon >0}\mathsf{\bar{G}}%
\left( \varepsilon \right) =\mathsf{G.}$

Let $N=N\left( \varepsilon \right) $ be so big, that the cube $B_N=\left\{ 
\mathbf{y}=\left( y_1,...,y_{d+1}\right) \in \Bbb{R}_{+}^{d+1}:0\le y_i\le
N\right\} $ contains the set $\mathsf{\bar{G}}\left( \varepsilon \right) 
\mathsf{.}$ We denote by $\mathbf{x}_N$ the vertex $\left( N,...,N\right) $
of this cube. Consider the convex set $\mathsf{\tilde{U}}=B_N\cap \mathsf{K}%
\left( \varepsilon \right) .$ We are going to define with its help a
function $T_N\left( \mathbf{n}\right) \equiv T_{N,\varepsilon }\left( 
\mathbf{n}\right) $ on $S^d.$ First, let $\mathbf{n}\in -\left( \Delta
^d\right) ;$ in other words, $\mathbf{n}$ has all coordinates non-positive.
Note that by definition the support plane $L^{=}\left( \mathbf{x}_N\mathbf{%
,n;}\tau _{\mathbf{x}_N,\mathsf{\tilde{U}}}\left( \mathbf{n}\right) \right) $
intersects the set $\mathsf{\bar{G}}\left( \varepsilon \right) .$ In case
when this intersection contains ``inner'' points of $\mathsf{\bar{G}}\left(
\varepsilon \right) ,$ i.e. points not in $\partial \mathsf{\bar{G}}\left(
\varepsilon \right) \equiv \mathsf{G}\cap \partial \left( \Bbb{R}%
_\varepsilon ^{d+1}\right) ,$ we put 
\begin{equation}
T_N\left( \mathbf{n}\right) =-\left( \mathbf{x}_N,\mathbf{n}\right) -\eta
\left( -\mathbf{n}\right) >0,  \label{22}
\end{equation}
where $\eta $ is our initial function (\ref{05}). We use the same definition
(\ref{22}) for remaining $\mathbf{n}$-s in $-\left( \Delta ^d\right) ,$ for
which the intersection 
\[
L^{=}\left( \mathbf{0,-n;}\eta \left( -\mathbf{n}\right) \right) \cap 
\mathsf{\bar{G}}\left( \varepsilon \right) \neq \emptyset . 
\]
For future use we denote the set of $\mathbf{n}$-s, where the function $T_N$
is already defined, by $\left( -\left( \Delta _\varepsilon ^d\right) \right)
;$ note that $\Delta _\varepsilon ^d\rightarrow \Delta ^d$ as $\varepsilon
\rightarrow 0.$ For the remaining $\mathbf{n}\in -\left( \Delta
^d\,\backslash \,\Delta _\varepsilon ^d\right) $ we define $T_N\left( 
\mathbf{n}\right) =\tau _{\mathbf{x}_N,\mathsf{\tilde{U}}}\left( \mathbf{n}%
\right) .$ For $\mathbf{n}$-s in $S^d\,\backslash \,\Delta ^d$ the function $%
T_N\left( \mathbf{n}\right) $ is defined by applying multiple reflections in
the coordinate planes. In other words, the values $T_N\left( \pm n_1,\pm
n_2,...,\pm n_{d+1}\right) $ do not depend on the choice of signs.
Analogously, we define the convex set $\mathsf{U}$ as the union of $\mathsf{%
\tilde{U}}$ and all its multiple reflections in coordinate planes shifted by 
$\mathbf{x}_N.$

It follows from the definitions above that the set $\mathsf{U}$ is nothing
else but the shift of the Wulff body $K_{T_N}^{<}\left( 1\right) $ by the
vector $\mathbf{x}_N.$ According to what was said in the section 2.2, for
every $M\in D_{\mathrm{vol}\left( \partial \mathsf{U}\right) },M\neq
\partial \mathsf{U},$ 
\begin{equation}
\mathcal{W}_{T_N}\left( \partial \mathsf{U}\right) <\mathcal{W}_{T_N}\left(
M\right) .  \label{23}
\end{equation}
Consider now an arbitrary hypersurface $\mathsf{H},$ such that $\partial 
\mathsf{H}=\partial \mathsf{\bar{G}}\left( \varepsilon \right) ,$ while the
set $\nu \left( \mathsf{H}\right) $ of its unit normal vectors belongs to
the subset $\Delta _\varepsilon ^d\subset \Delta ^d$ (which is the case for
the surface $\mathsf{\bar{G}}\left( \varepsilon \right) $ itself). Then for
any such $\mathsf{H}$ 
\begin{equation}
\mathcal{V}_\eta \left( \mathsf{H}\right) +\mathcal{W}_{T_N}\left( \mathsf{H}%
\right) =N\sqrt{d}\,\mathrm{vol}\left( \pi \left( \partial \mathsf{\bar{G}}%
\left( \varepsilon \right) \right) \right) ,  \label{21}
\end{equation}
where $\pi \left( \partial \mathsf{\bar{G}}\left( \varepsilon \right)
\right) $ is the projection of the ``curve'' $\partial \mathsf{\bar{G}}%
\left( \varepsilon \right) $ (of codimension 2) on the hyperplane $\left\{ 
\mathbf{y}:y_1+...+y_{d+1}=0\right\} \subset \Bbb{R}_{+}^{d+1},$ and where $%
\mathrm{vol}\left( \pi \left( \partial \mathsf{\bar{G}}\left( \varepsilon
\right) \right) \right) $ is the ($\left( d-1\right) $-dimensional) volume
inside it. The relation (\ref{21}) follows from (\ref{22}). Therefore the
minimality property (\ref{23}) of the functional $\mathcal{W}_{T_N}$ on the
surface $\mathsf{\bar{G}}\left( \varepsilon \right) $ implies the maximality
property of the functional $\mathcal{V}_\eta $ on the same surface!

The uniqueness statement for $\mathcal{V}_\eta $ is therefore a corollary of
the uniqueness for $\mathcal{W}.\;$

\smallskip It remains now to consider the question when the volumes $\mathrm{%
vol}\left( G_\eta \left( \lambda \right) \right) $ are infinite for all $%
\lambda .$ We are going to show that in that case $v_\eta =\infty .$ To make
things look simpler, we restrict ourselves to the 2D case. Let $G\subset
G_\eta \left( 1\right) $ be an arc, and consider the ``triangle'' $\Delta
\left( G\right) \subset \Bbb{R}^2,$ made from all the points of all the
segments joining the origin to the curve $G:$ 
\[
\Delta \left( G\right) =\bigcup_{x\in G}\left[ 0,x\right] . 
\]
It is straightforward to see that 
\[
\mathrm{vol}\left( \Delta \left( G\right) \right) =\frac 12\mathcal{V}_\eta
\left( G\right) . 
\]
We now will present the family $G_\gamma \in \bar{D}_1,$ such that $\mathcal{%
V}_\eta \left( G_\gamma \right) \rightarrow \infty $ as $\gamma \rightarrow
0.$ Namely, for every $\lambda $ we define the number $N\left( \lambda
\right) $ to be the size of the square $B\left( \lambda \right) =\left\{ 
\mathbf{y}\in \Bbb{R}^2:0\le y_i\le N\left( \lambda \right) \right\} $ for
which $\mathrm{vol}\left( Q\left( G_\eta \left( \lambda \right) \right) \cap
B\left( \lambda \right) \right) =1,$ and we put $G_\gamma $ to be the part
of the boundary of the intersection $Q\left( G_\eta \left( \gamma \right)
\right) \cap B\left( \gamma \right) ,$ which is visible from the point $%
\left( 2N\left( \gamma \right) ,2N\left( \gamma \right) \right) ,$ say. The
curve $G_\gamma $ consists of a certain arc $\bar{G}_\gamma $ of the curve $%
G_\eta \left( \gamma \right) $ and two small segments, joining its endpoints
to the coordinate axes. By construction, $\mathrm{vol}\left( \Delta \left( 
\bar{G}_\gamma \right) \right) >\frac 13.$ On the other hand, $\mathrm{vol}%
\left( \Delta \left( \bar{G}_\gamma \right) \right) =\frac \gamma 2\mathcal{V%
}_\eta \left( \bar{G}_\gamma \right) ,$ which implies that 
\[
\mathcal{V}_\eta \left( \bar{G}_\gamma \right) >\frac 2{3\gamma }. 
\]

\endproof%

\section{Conclusion}

In this paper we have described the explicit geometric construction, which
predicts the asymptotic shape of some combinatorial objects. It is worth
mentioning that the method presented should work whenever the underlying
probability measure has certain locality property, namely that the distant
portions of the combinatorial object under consideration are weakly
dependent. This locality property is in fact the key feature behind the
results obtained in the papers cited above. It also holds for the
corresponding problems of statistical mechanics, like the validity of the
Wulff construction, and is crucial there as well.\medskip

\textbf{Acknowledgment.\ }I would like to thank the referees for their
valuable remarks.


\begin{thebibliography}{ACC}
\bibitem[ACC]{ACC}  K. Alexander, J.L. Chayes and L. Chayes: \textit{\ The
Wulff construction and the asymptotics of the finite cluster distribution
for the two-dimensional Bernoulli percolation,} Comm. Math. Phys., \textbf{%
131}, 1-50, 1990.

\bibitem[Bl]{Bl}  A. Blinovskij. \textit{LDP for shape of random Young
diagram}, to appear in: Information Transmission Problems, 1999.

\bibitem[Bo]{Bo}  T. Bodineau. \textit{The Wulff construction in three and
more dimensions}, Preprint 1999.

\bibitem[Ce]{Ce}  R. Cerf, \textit{Large deviations for three dimensional
supercritical percolation}, Preprint 1998.

\bibitem[CeP]{CeP}  R. Cerf and A. Pizstora. \textit{On the Wulff crystal in
the Ising model.} Preprint, 1999.

\bibitem[CKP]{CKP}  H. Cohn, R. Kenyon and J. Propp. \textit{A variational
principle for domino tilings.} Preprint, 1998.

\bibitem[DVZ]{DVZ}  A. Dembo, A. Vershik and O. Zeitouni. \textit{Large
deviations for integer partitions.} Preprint, 1998.

\bibitem[DKS]{DKS}  R.L. Dobrushin, R. Kotecky and S.B. Shlosman: \textit{%
Wulff construction: a global shape from local interaction, }AMS translations
series, Providence (Rhode Island), 1992.

\bibitem[I1]{I1}  D. Ioffe. \textit{Large deviations for the 2D Ising model:
a lower bound without cluster expansions,} J.Stat.Phys., \textbf{74},
411-432, 1994.

\bibitem[I2]{I2}  D. Ioffe. \textit{Exact deviation bounds up to }$T_c$%
\textit{\ for the Ising model in two dimensions,} Prob. Th.. Rel. Fields 
\textbf{102}, 313-330, 1995.

\bibitem[IS]{IS}  D. Ioffe and R.Schonmann. \textit{%
Dobrushin-Kotecky-Shlosman theory up to the critical temperature,} Comm.
Math. Phys., \textbf{199}, 117-167, 1998.

\bibitem[Ke]{Ke}  R. Kenyon. \textit{The planar dimer model with boundary: a
survey. }Preprint, 1998.

\bibitem[T1]{T1}  J.E. Taylor, \textit{Unique structure of solutions to a
class of nonelliptic variational problems, }Proc. Symp. Pure Math., \textbf{%
27}, 419-427, 1975.

\bibitem[T2]{T2}  J.E. Taylor, \textit{Some crystalline variational
techniques and results, }Asterisque, \textbf{154-155}, 307-320, 1987.

\bibitem[V1]{V1}  A. Vershik, \textit{Statistical mechanics of combinatorial
partitions, and their limit configurations.} (Russian) Funktsional. Anal. i
Prilozhen. \textbf{30}, no. 2, 19--39, 1996; translation in Funct. Anal.
Appl. \textbf{30,} no. 2, 90--105, 1996.

\bibitem[V2]{V2}  A. Vershik. \textit{Limit distribution of energy of
quantum ideal gas from the point of view of the theory of partitions of
natural numbers}, Russ. Math. Surv., \textbf{52}, 139-146, 1997.

\bibitem[VK]{VK}  A. Vershik and S. Kerov. \textit{Asymptotic of the largest
and typical dimensions of irreducible representations of a symmetric group.}
Funct. Anal. Appl., \textbf{19}, 21-31, 1985.

\bibitem[W]{W}  G. Wulff, \textit{Zur frage der geschwindigkeit des
wachsturms under auflosung der kristallflachen, }Z.Kristallogr. \textbf{34},
449-530, 1901.
\end{thebibliography}
\end{document}